\documentclass[aps,prb,twocolumn,footinbib,showpacs,floatfix]{revtex4}
\newcommand{\bel}[1]{  \begin{equation} \label{#1}}
\newcommand{\eel}{\end{equation}}

\newcommand{\fra}[2]{\displaystyle \frac{#1}{#2}}
\newcommand{\Traza}[1]{\mbox{Tr}\,\left\{#1\right\}}
\newcommand{\ket}[1]{\left| #1 \right>}
\newcommand{\bra}[1]{\left< #1 \right|}

\newcommand{\ketup}{\ket{\uparrow}}
\newcommand{\ketdown}{\ket{\downarrow}}
\newcommand{\braup}{\bra{\uparrow}}
\newcommand{\bradown}{\bra{\downarrow}}

\newcommand{\der}[2]{\displaystyle \frac{\mbox{d}\;\!#1}{\mbox{d}\;\!#2}} 
\newcommand{\dd}{\mbox{d}}
\usepackage{amssymb}
\usepackage[pdftex]{graphicx}	
	\DeclareGraphicsExtensions{.pdf,.png,.jpg}
\usepackage[applemac]{inputenc}
\usepackage{amsmath,amssymb,amsthm}
\usepackage{mathbbol,bbm}
\usepackage{graphicx,float}
\usepackage[final]{showkeys}
\usepackage{bm,dsfont}

\def\ket#1{\left|#1\right>}
\def\bra#1{\left<#1\right|}

\def\dd{\mathord{\rm d}}

\usepackage{booktabs}

\usepackage{xcolor}
\usepackage{amsfonts}
\usepackage{tikz}
\usetikzlibrary{calc}
\usetikzlibrary{trees}
\usetikzlibrary{snakes}
\usetikzlibrary{automata}
\usetikzlibrary{arrows}
\usetikzlibrary{positioning}
\newcommand{\OSC}{$\rm DMC$\,}
\newcommand{\CT}{$\rm CT$\,}
\newcommand{\eigrho}{{\mathfrak{r}}}

\begin{document}
\title{Quantum Information Engines with Many-Body States \\
attaining Optimal Extractable Work with Quantum Control}
\author{J.M. Diaz de la Cruz$^{1}$ and M.A. Martin-Delgado$^{2}$}
\affiliation{$^1$Departmento de Fisica Aplicada, Universidad Politecnica, 28006 Madrid, Spain  \\ $^2$Departamento de Fisica Teorica I, Universidad Complutense, 28040 Madrid, Spain}

\vspace{-3.5cm}

\begin{abstract}
We introduce quantum information engines that extract work from quantum states and a single thermal reservoir.  They may operate under three general conditions: i/  Unitarily Steered evolution (US), driven by a restricted set of available  Hamiltonians; ii/ Irreversible Thermalization (IT) and iii/ Isothermal Relaxation (IR), and hence are called USITIR machines. They include novel engines without traditional feedback control mechanisms, as well as versions which also include them. Explicit constructions of USITIR engines are presented for one- and two-qubit states and  their maximum extractable work is computed, which is optimal. Optimality is achieved when the notions of controllable thermalizability and density matrix controllability are fullfilled. Then, many-body extensions of USITIR engines are also analyzed and  conditions for optimal work extraction are identified. When they   are not met, we  measure their lack of optimality by means of  newly defined {\em uncontrollable entropies,} that are explicitly computed  for some selected examples. This includes  cases of distinguishable and indistinguishable particles. 
\end{abstract}




\maketitle

\section{Introduction}
\label{sec:intro}


The goal of increasing the speed of a universal computer has played 
a central role in computer design, but this comes along with a big power bill
that also scales up the faster the  computer is. Thus, power saving has become a
central issue in the new generations of supercluster computers. This has led to
the creation of the Green500 list \cite{green500} with the world's most energy-efficient supercomputers 
that complements the traditional TOP500 list \cite{top500} ranking the most
powerful non-distributed computers. The possibility of doing classical universal computation in
a fully reversible way \cite{fredkin_toffoli82} was motivated by the quest of a dissipationless computer 
\cite{bennett73, keyes_landauer70, likharev82, landauer61}.
 Interestingly enough, the early origins of quantum 
computing are linked to the research on the energy requirements of standard computers
since reversibility  was  considered as a resource to reduce energy dissipation \cite{rmp}. A first hint
to the idea of a quantum computer was to use the unitary evolution of quantum mechanics
as a natural way to achieve the goal of reversibility in computation \cite{benioff81, benioff82}.  Yet, this missed
the new possibilities offered by quantum superposition and entanglement properties of
quantum states.
However, soon thereafter the surprising new capabilities offered by quantum computers 
\cite{feynman82, deutsch85}, with speedups over classical algorithms, took over the original energy considerations and has been the predominant topic in the theory of quantum computation. The experimental feasibility of realizing a real quantum computer \cite{cirac_zoller95} has been another essential breakthrough for 
the development of quantum computation as a new field of research.

Soon after the model of universal quantum computer became accepted as a new model 
for computation, it was evident that a realistic experimental realization of it  could only be possible by addressing again the problems of dissipation, this time  in the context of battling the decoherence   disturbing effects \cite{shor95, steane96} that inexorably affect  the entangling properties of quantum states needed to carry out a powerful quantum computation. This way, thermodynamical aspects of quantum computation are making a come back. A quantum computer can be thought of as a special type of engine that trades free energy for mathematical operations at the expense of dissipating some heat. In this work, we focus on another type of machines known as quantum information engines whose main goal is to extract the  maximum amount of work from possibly entangled many-body states regardless of the computational work that they may carry out. Quantum engines play a fundamental role in investigating quantum effects in thermodynamical laws, in particular the second law.

An essential question in Thermodynamics is what are the limits for the work we can extract from a given system, be it classical or quantum.  The traditional machine used to address this question is the Szilard engine (SZE)\cite{SZE}. The classical case of the SZE was solved by Bennett \cite{bennett82} realizing that the second law of thermodynamics can be recovered once we take into account the dissipation introduced in the erasure of the register needed to reset the thermodynamic cycle of the engine.  The necessity of this energy cost dissipated as heat is called the Landauer principle \cite{landauer61}. 
In the quantum world, several studies have been carried out with quantum versions of the SZE \cite{lloyd97, scully01, zurek03, scully05, rostovtsev05, sariyanni05, QWJSN06, sagawa_ueda09, RARDV11, KSDU11, PDGV12, FWU12, PKSK13, Deliberato2011}. Contrary to the classical case, the energy analysis in the quantum engine is more involved,  since the
role of the piston or barrier present in the SZE introduces energy costs in several stages of the engine cycle that are not present in the classical engine. Moreover, the quantum SZE operates with a feedback control mechanism \cite{KSDU11, PKSK13} that is an essential ingredient to extract work from quantum information.


In this paper we use different variations of a relatively simple quantum information heat engine based on magnetic qubits\cite{lloyd97, piechocinska2000} as concept models to process information contained in multi-qubit systems. We show that the premises for optimality of the extractable work refer to quantum control theory\cite{Dalessandro2007} and to a problem that we call controlable thermalizability, put forward in section IV. It is shown that the latter implies a specially demanding condition when many-body quantum states are used. Consideration for restricted sets of available Hamiltonians follows from this difficulty.

 Under non-optimal conditions, we  also determine the relation between the obtainable work and a new variation of relative entropy that we call  {uncontrollable entropy}. 
The  quantum information heat engine (QIHE) that we propose avoids the use of a barrier as a basic component in the quantum engine. Information Heat Engines are devices that extract work from a thermal reservoir in exchange for an increment of entropy in some physical system at the core of the machine. Cyclic workflows imply procedures to reset the entropy of such system. The best known way for it is to purify the quantum state of the internal system by performing a Von Neumann measurement, at the cost of increasing the entropy of the qubits that hold the result. A less visited way is to swap the entropy of internal and fresh quantum systems that are cyclically fed to the machine. Considering the connection between information theoretic concepts and thermodynamic entropies, some papers\cite{PKSK13,FWU12} have explored the extraction of work from the entanglement of a two-qubit system. We analyze a many-body generalization and conclude that there is a non trivial issue concerning the set of hamiltonians that drive the evolution of the system. The new QIHEs are introduced for the simple cases using one and two qubit systems. Feedback control versions of them
are also constructed. In both cases, the extractable work analysis shows that it is maximal for these quantum engines. As this maximum work coincides with the one compatible with the second law of thermodynamics, then it is optimal. We generalize these QIHEs based on swap operations to systems comprising many-body states and perform the extractable work analysis yielding the optimal value. 
We call them USITIR machines, after the three kinds of evolution under which they may operate: Unitary Steering, Irreversible Thermalization and Isothermal Relaxation. This generalization includes  novel models as well as previously well-known models of quantum engines.
We notice that in order to achieve optimality for these USITIR engines with  more than 2 qubits,  general $k$-body control Hamiltonians are needed.

In a quantum many-body system, there is the issue of whether the particles are identical \cite{KSDU11} or not. Thus, for the many-body USITIR engines we also derive the corresponding formulas of the optimal extractable work for identical particles \cite{KSDU11, PDGV12}.

This paper is organized as follows: in section \ref{sec:qengine}, we present  simple and scalable models of Magnetic Information Heat Engines, which motivate the issues  of controllability and controllable thermallizability developed in sections \ref{sec:qcontrol} and \ref{sec:qthermo}, respectively. These properties prompt for a more abstract and general type of Information Heat Engine that is introduced in section \ref{sec:qmany}. We refer to them as USITIR machines  and the work that they can process is analyzed in section \ref{sec:qwork}. Section \ref{sec:qconclusions} is devoted to conclusions. In several appendices we provide detailed calculations for the  uncontrollable entropies and extractable work of some relevant USITIR engines.

An extensive study on the rich relations between relative entropy and thermodynamics has been published \cite{Vedral2002} and also from a resource perspective \cite{Oppenheim2013, Brandao2013}. Recently a number of publications extends the validity of relations between work and entropy by introducing generalizations of the Shannon or Von Neumann entropy to the so called {\em smooth} entropies. They can handle situations where the goal is not merely maximizing the expected value of the extracted work, but also other aspects in regard of its fluctuations and probabilities of failure\cite{Dahlsten2011,Horodecki2011,Egloff2013}. However,  our paper only considers machines that work in cycles and focuses only on the expected value of the extracted work.

\section{Quantum Engines based on Magnetic Qubits}
\label{sec:qengine}
This section presents an alternative to previous models of Quantum Information Heat Engines\cite{KSDU11,Toyabe2010,Zhou2010}. In the following paragraphs we describe a Magnetic Quantum Information Heat Engine (henceforth MQIHE) both for 1-qubit and 2-qubit inputs  which is different from other magnetic alternatives\cite{lloyd97,piechocinska2000}. Unlike most previously described Szilard Engines, some of our models contain neither an {explicit} measurement nor a feedback control. It is simpler than the previously described Information Heat Engines and shows clearly how it serves its purpose: to trade entropy for work, taking energy from a heat reservoir at a definite temperature using up the information provided by input qubits. We describe two kinds of devices: swap and feedback engines. In swap machines, the density matrix of the input system (one qubit in subsection \ref{subsec:one-qengine} or two qubits in subsection \ref{subsec:two-qengine}) and that of internal magnetic qubits are swapped. In feedback engines, the internal qubits are entangled with the input ones (one qubit in subsection \ref{subsec:feedback} or two qubits in subsection \ref{subsec:two-feedback}) which are subsequently measured and the outcomes govern further action. The degree of purity of the input qubits determines the quality of the measurement and thus the extractable work. We also provide physical representations of the engines.
\subsection{One-qubit Magnetic Quantum Engine}
\label{subsec:one-qengine}
\begin{figure}
\includegraphics[width=0.35\textwidth]{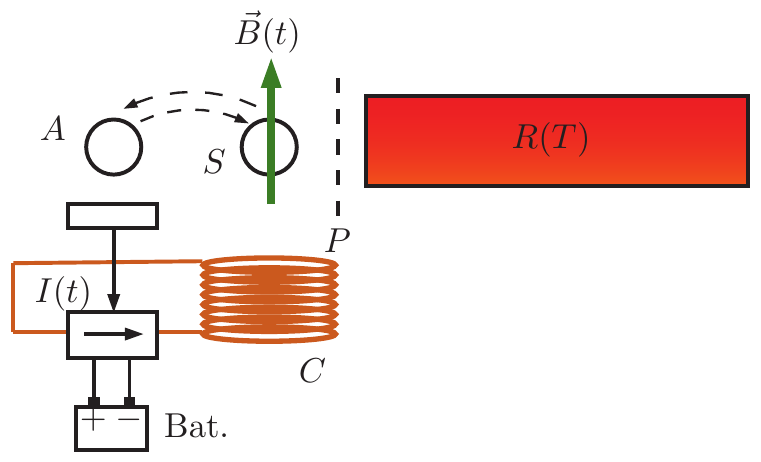}
\caption{(Color online) Main components of the 1MQIHE: the information of input qubit $A$ is traded for that of a magnetic qubit spin$\frac 1 2$ $S$ that lies in the magnetic induction $\vec B(t)$ created by the current $I(t)$ running through coil $C$. The power needed to drive the current is taken from/to an electrical battery. The qubit $S$ may be set in thermal equilibrium with or isolated from a heat bath $R(T)$ at fixed temperature $T$.}
\label{fig::motor}
\end{figure}

  In this subsection we describe a one-qubit MQIHE, hence 1MQIHE, showing how, after each cycle, some energy is stored in an electrical battery provided that the system is fed with a non fully depolarized magnetic qubit; this qubit exits the cycle completely depolarized. The system depicted in Fig. \ref{fig::motor} represents a 1MQIHE.

There are two important qubits in the system: the {\em ancilla} or {\em input} $A$ and the {\em system} or {\em internal} qubit $S$. The ancilla $A$ is an external qubit that enters the machine in a known, possibly mixed, state; it will be shown that each ancilla in a pure state will enable us to obtain an energy  $k_B \,(\ln 2)\,T$ from the thermal reservoir $R(T)$ which is always supposed to be in equilibrium at temperature $T$. The $S$ qubit is a magnetic spin $\frac 1 2$ whose state may belong to the Hilbert space spanned by the kets $\ketup,\ketdown$ or may even be a mixed state without further restriction. The magnetic qubit $S$ lies in a magnetic field  $\vec B(t)=B(t)\,\vec u_z$ generated by the electrical current $I_C(t)$ running through coil $C$. The thermal connection between the $S$ qubit and the reservoir $R(T)$ is accomplished through a wall $P$ which may be adiabatic or diathermal, according to a cyclic timing that is described below. 

In order to appreciate more clearly the trade-off between information and energy, we assume that the energies of the $S$ and $A$ qubits are completely degenerate in the initial and the final instants. This means that the energy  of the qubit $A$ is the same irrespective of whether it is in the $\ketup,\ketdown$ state or any mixture thereof. This is also true for qubit $S$.

Next, a description of the three stages of the 1MQIHE cycle is given along with the evolution of the states of the $A,S$ qubits; the energy flow is traced in appendix \ref{sec:energ} to draw a final balance. 

\begin{itemize}
\item[i) ]At this stage, the ancilla $A$ and the system qubit $S$ stay thermally isolated from the reservoir $R(T)$ and undergo a reversible swap operation whereby  $\forall \rho_p,\rho_q\,,\,(\rho_p)_A \otimes (\rho_q)_S$ transforms into $(\rho_q)_A \otimes (\rho_p)_S$. It is important to remark that the swapping is not physical, but just logical. A possible physical way to implement a SWAP gate between magnetic qubits is by means of the Heisenberg exchange interaction which performs this operation in a reversible way .

The $S$ qubit starts the cycle as it finishes the last stage, so that it is in equilibrium with the thermal reservoir in a completely depolarized state $\rho_{S,0}=\frac 1 2 {\mathbbm 1}$, with no applied magnetic field; the ancilla begins in a known, possibly mixed, state $\rho_{A,0}$. After the swapping, the ancilla is released in a completely depolarized state $\rho_{A,1}=\frac 1 2 {\mathbbm 1}$ and the system qubit $S$ is left in the $\rho_{A,0}$ state. Now an auxiliary magnetic field is applied to rotate the $\rho_A$ state in order to align its spin with the axis of the coil (the $z$ axis). This means that the $S$ qubit exits this stage in a state
\begin{equation}
\rho_{S,1} = \frac 1 2 \left( \mathbbm{1} + c \sigma_z \right),
\end{equation}
with the same entropy as the initial ancilla:  $S(\rho_{A,0})=S(\rho_{S,1})$; moreover, we can assume, without loss of generality, that $c\geq 0$.

\begin{figure}
\includegraphics[width=0.4\textwidth]{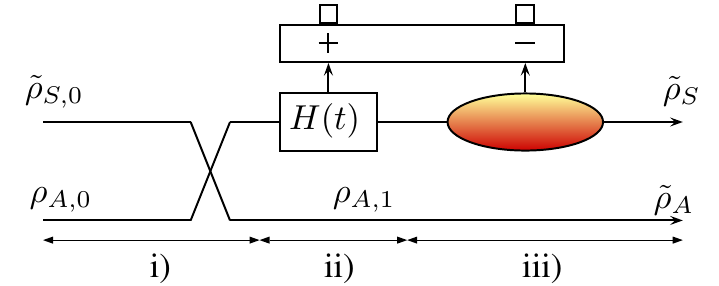}
\caption{(Color online) Simplified workflow of 1MQIHE. Density matrices of qubits $A,S$ are swapped; then a time varying current $I_C(t)$ determines a Hamiltonian $H(t)$ which acts  isolated from $R(T)$ at stage ii) and in thermal contact at stage iii). The shaded oval means evolution under thermal equilibrium with $R(T)$ and $\tilde{\rho}_S$ represents a completely depolarized state for qubit $S$.}
\label{fig::esmotor}
\end{figure}

So far, no energy has been interchanged with the coil. 
\item[ii) ] Next, $S$ keeps thermally isolated from $R(T)$, the current source is turned on and the current through the coil $I_C$ grows from zero to a maximum value $I_M$. The magnetic field created by $I_C$, which is in the $z$ direction, grows accordingly. At this stage, the system qubit does not change, because its state commutes with the time varying Hamiltonian, therefore the magnetic moment is also constant $\mu_{z,0}$.  The final value of $I_C$ corresponds to a value $B_f$ of $B$ determined under the condition that the overall energy extraction is maximal. It is shown in Appendix \ref{sec:energ} that the optimum value of $B_f$ is the one that makes the $S$ qubit state $\rho_{S,1}$ to be in equilibrium with the reservoir at temperature $T$.

\item[iii) ] Finally, the system qubit $S$ is put in thermal contact with the reservoir and the current $I_C$ is gradually lowered so that the qubit keeps in equilibrium with the reservoir. In this stage the current source recovers energy in excess of the one previously supplied. The equilibrium for $\rho_S$ when the current is set to zero is the completely depolarized state and the engine is ready for the next cycle.
\end{itemize}


\begin{figure}
\includegraphics[width=0.42\textwidth]{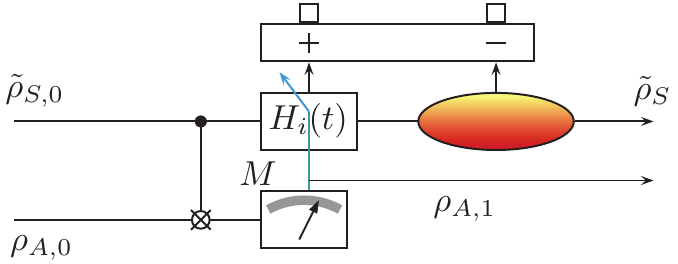}
\caption{(Color online) Simplified workflow of the feedback version of 1MQIHE (see Fig.\ref{fig::esmotor}). The CNOT(S,A) gate is followed by a measurement of $A$; its result determines the choice of $H_i(t)$; then it proceeds like MQIHE.}
\label{fig::esmotorb}
\end{figure}

\begin{figure}[t]
\includegraphics[width=0.5\textwidth]{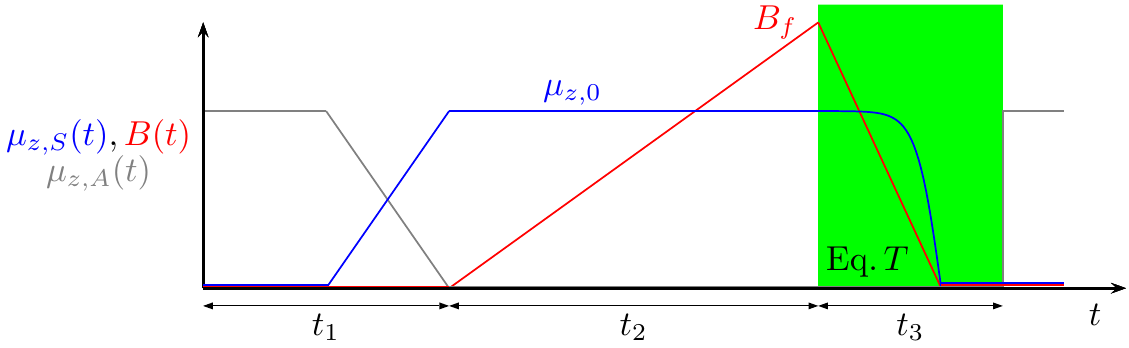}
\caption{(Color online) Evolution of the magnetic field $B(t)$ (red) and the $z$ components of the magnetic moments $\mu_z$ of qubits $A$ (grey) and $S$ (blue). The green shaded area corresponds to thermal equilibrium with the reservoir.}
\label{fig::energiaj}
\end{figure}

The thermal contact is interrupted when the cycle starts once more until the stage iii) is reached again. Fig. \ref{fig::energiaj} shows the evolution of $B(t)$, $\mu_z(t)$ and $R(t)$ with all magnitudes normalized.


We claim that 1MQIHE is optimal, in the sense that it obtains the maximum theoretically allowable work from the processing of the qubit $A$, in the presence of a thermal reservoir at temperature $T$. It is well known\cite{sagawa_ueda09,Sagawa2012}, that  the maximum amount of work that a system {\cal S} can obtain from a bath is given by 
\begin{equation}
W_{1\to 2} =-\frac{S(\rho_1||\rho_\beta(H_1))-S(\rho_2||\rho_\beta(H_2))}{\beta} \ln 2 -\Delta F,
\end{equation}
where $\rho_1,\rho_2,H_1,H_2$ are the initial and final states and Hamiltonians of qubit {\cal S}, $\rho_\beta(H)$ is the state that corresponds to thermal equilibrium with the reservoir for the Hamiltonian $H$, $\Delta F$ is the difference  of the free energies of states $\rho_\beta(H_1),\rho_\beta(H_2)$ and $S(\sigma||\tau)=\Traza{\sigma\ln\sigma-\sigma\ln \tau}$ is the relative entropy of $\sigma$ with respect to $\tau$. If $H_2=H_1=0$, then $
W_{1\to 2} =-\frac{S(\rho_1||\rho_\beta(H_1))-S(\rho_2||\rho_\beta(H_2))}{\beta} \,\ln 2
$
and if we also assume that the final state is in thermal equilibrium and the dimension of the Hilbert space is $d$,we have $W_{1\to 2} =\frac{\log_2 d-S(\rho_1)}{\beta} \ln 2$ which, if the system is a set of $N$ qubits, yields
\bel{opti}
W_{1\to 2} =\frac{N-S(\rho_1)}{\beta} \ln 2.
\eel
A detailed deduction of the work obtained in the cycle is presented in appendix \ref{sec:energ}. The result is eq.\eqref{es} that represents the particularization of eq.\eqref{opti} for one qubit. 


Two problems emerge from the previous consideration: \begin{enumerate}
\item[a)] is there a thermal equilibrium state unitarily equivalent to an arbitrary initial state?
\item[b)] can the system be unitarily steered from an arbitrary initial state to thermal equilibrium? 
\end{enumerate}

In subsection \ref{subsec:two-qengine} we extend the 1-qubit MQIHE to a two-qubit MQIHE, answer the previous questions and then show that with a controllable Heisenberg interaction the machine is still optimal. In subsection \ref{subsec:feedback} we describe a feedback model for the 1MQIHE.
\subsection{Feedback version of 1MQIHE}
\label{subsec:feedback}
A feedback version of 1MQIHE (Fig. \ref{fig::esmotorb}) has a slightly different cycle.  At stage i), instead of logically swapping qubits $A$ and $S$, a CNOT$(S,A)$ gate, controlled by $S$, is applied. We assume that $\rho_{A,0}=\frac 1 2 \left( \mathbbm{1} + c \sigma_z \right)$; if it is not, an auxilliary field is used to rotate the state.   After the CNOT$(S,A)$ stage, the bipartite state is:
\begin{eqnarray}
\rho_{AS,1}=&\frac 1 4 &\left[(1+c)\ket{\uparrow \, \uparrow}_{AS}\bra{\uparrow \, \uparrow} + (1-c)\ket{\downarrow \, \uparrow}_{AS}\bra{\downarrow \, \uparrow} +\right.\nonumber\\
&+& (1+c)\ket{\downarrow \, \downarrow}_{AS}\bra{\downarrow \, \downarrow} + \left.(1-c)\ket{\uparrow \, \downarrow}_{AS}\bra{\uparrow \, \downarrow} \right].\nonumber\\
&\,& \,
\end{eqnarray}

Next, qubit $A$ is measured in the computational basis; the result $p$ may be either $p=1$ or $p=2$, corresponding to $\ketup$ or $\ketdown$, respectively and the post-measurement state $\rho_{i,p}$ of $S$ is
\begin{equation}
\rho_{i,p} = \frac 1 2 \left(
{\mathbbm 1}_S \,\pm \,c \,\sigma_{z,S}
\right),
\end{equation} 
where the plus or minus signs hold for $p=1$ or $p=2$, respectively. In the first case ($\ket{A}=\ketup$), the $S$ qubit is in the same state as in the end of stage ii) for 1MQIHE and the machine proceeds identically, otherwise ($\ketdown$) the same analysis applies provided that the coil current and, accordingly, the magnetic flux density $B$, are inverted. Consequently, the energy balance is also the same.

\vspace{2mm}

\subsection{Two-qubit Magnetic Quantum Engine}
\label{subsec:two-qengine}

Information stored in single qubits can be used to optimally extract work, as it has been proved in the two previous subsections. However,  information can also be stored in two-qubit systems that may be entangled. In section \ref{sec:qwork} it is shown that one-qubit control Hamiltonians, e.g. Zeeman interactions with two external magnetic fields, cannot extract the optimal amount of work from a general, possibly entangled, two-qubit system. In order to open the generalization of Information Heat Engines to many-qubits  we now define a system endowed with a well-known interaction: the Heisenberg two-spin interaction acting on a system consisting of two magnetic spins $\frac 1 2$. 

The two-qubit Magnetic Quantum Information Heat Engine (2MQUIHE, henceforth) contains two system qubits $S_1,S_2$ and two ancillas $A_1,A_2$. At the beginning of each cycle $S_1,S_2$ are completely depolarized, whereas qubits $A_1,A_2$ enter in an arbitrary, possibly mixed, state defined by the $4\times 4$ density matrix $\rho_i$. At the end of each cycle, all four qubits exit in a completely depolarized and hence disentangled state. Work is extracted from the increase of entropy of the two-qubit system $A_1,A_2$. It is supposed that the Hamiltonian of qubits $A_1,A_2,S_1,S_2$ is the same and completely degenerate at the beginning and  the end of each cycle.

\begin{figure}
\begin{center}
\hspace{-8mm} \includegraphics[scale=1]{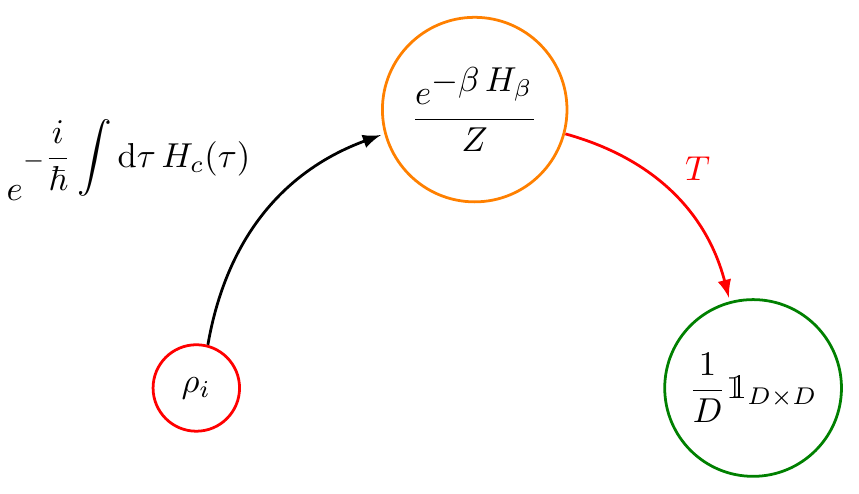}
\end{center}
\caption{(Color online) Schematic view of the transformation from the input state $\rho_i$ of the input qubits to a completely depolarized state. After transferring the state $\rho_i$ to the internal qubits, they should be driven to a thermal equilibrium state $\rho_\beta(H_\beta)$ by adequately steering the system through the Hamiltonian $H_c(t)$. This process corresponds to the left arrow line. The existence of an available Hamiltonian $H_\beta$ for which the equilibrium state $\rho_\beta(H_\beta)$ is unitarily reachable from $\rho_i$ through the control Hamiltonians is a central point of this paper. The right arrow line represents how, by gradually turning off the control Hamiltonians, the state becomes completely depolarized. }
\label{fig:flow}
\end{figure}

The states of qubits $S_1,S_2$ are initially swapped with those of $A_1,A_2$. Qubits $A_1,A_2$ now exit in a completely depolarized state. Then the $S_1,S_2$ pair is unitarily driven by an adequate steering of the magnetic fields $\vec B_1,\vec B_2$ acting individually on $S_1,S_2$ and of the distance between $S_1,S_2$ (tuning the strength of the Heisenberg interaction) to a state $\rho_\beta(H_\beta)$ in thermal equilibrium at temperature $T$ for the Hamiltonian $H_\beta$. In section \ref{sec:qthermo} the existence of the Hamiltonian $H_\beta$, and hence of the state $\rho_\beta(H_\beta)$, is proved and in section \ref{sec:qcontrol} the possibility of unitarily driving the system from any initial state to $\rho_\beta(H_\beta)$ is established.  Figure \ref{fig:flow} shows schematically the two stages. In the first one (black arrow), the initial state $\rho_i$ of $A_1,A_2$ which has been transferred to $S_1,S_2$ undergoes an unitary evolution driven by the Hamiltonian $H_c(t)$ which results from a suitable combination of the control Hamiltonians, as explained further in section \ref{sec:qcontrol}. The resulting state should be in thermal equilibrium at temperature $T$ for one available Hamiltonian $H_\beta$. The second stage is represented by the right arrow line and takes place entirely in thermal equilibrium at temperature $T$; the Hamiltonians are progressively switched off leading to a completely depolarized state at the end.

Finally, the steering fields are gradually turned off and the positions of $S_1,S_2$ are restored to their initial locations, extracting a work given by $-\Delta F$ where $F$ is the Helmhotz free energy of the system of qubits $S_1,S_2$. Under a completely degenerate Hamiltonian and being in thermal equilibrium,  qubits $S_1,S_2$ relax to a completely depolarized state.

An energy analysis equivalent to the one performed in subsection \ref{subsec:one-qengine} yields a work extraction given by
\begin{equation}
W=k_B T \,(\ln 2)\,(2-S(\rho_i)),
\end{equation}
which is the optimal value consistent with the second law of thermodynamics and the information content of the processed qubits $A_1,A_2$.

\begin{figure}[t]
\includegraphics[width=0.4\textwidth]{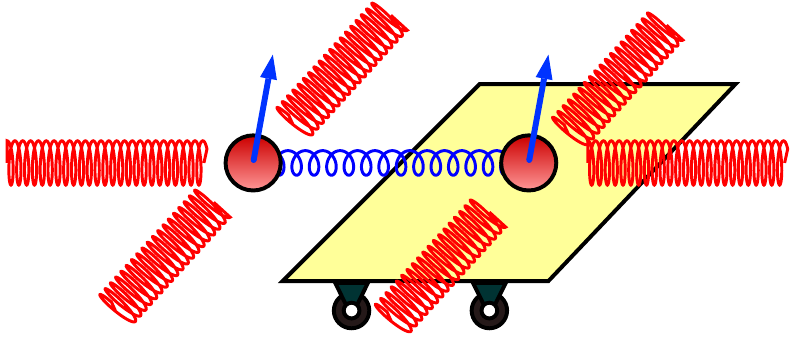}
\caption{(Color online) 2MQIHE engine outline: two possibly entangled qubits (joined blue line) lie in the magnetic fields created by a set of control currents running through the corresponding coils. The system evolves under the action of independent magnetic fields and a Heisenberg spin interaction suitably tuned by changing the distance.}
\label{fig::two-mqihe}
\end{figure}
\subsection{Feedback version of 2MQIHE}\label{subsec:two-feedback}
A feedback version of 2MQIHE follows immediately: instead of swapping qubits $A_1,A_2$ and $S_1,S_2$, the $CNOT(S_1,A_1),CNOT(S_2,A_2)$ are followed by a measurement of qubits $A_1,A_2$ on the computational basis. Because $S_1,S_2$ are completely depolarized before the operation, all four possible outcomes are equally likely, implying that $A_1,A_2$ exit completely unpolarized. After measurement, qubits $S_1,S_2$ are in the initial state $\rho_i$ of the ancillas $A_1,A_2$, except for a possible $NOT$-gate operation on each of the qubits, depending on the result of the measurement. Thus, all the controllability and thermalizability considerations previously written for the 2MQIHE also hold for its feedback version.

\section{Control Hamiltonians}
\label{sec:qcontrol}
In order to avoid  inefficient irreversible steps, Information Heat Engines should be able to maneuver the internal qubits in such a way that their density matrix undergoes no discontinuity when thermal equilibrium with the reservoir sets in. It is then imperative to provide an effective control system able to drive the internal qubits from the input state to another one which is in thermal equilibrium at temperature $T$. The process is depicted in Fig.\ref{fig:flow}. 

In this section we stick to a typical quantum control scenario where a system with density matrix $\rho$ evolves unitarily under a Hamiltonian given by a linear combination $H_c=\sum_{j=1}^C c_j(t) H_j$ where ${\cal C}=\{H_j\}$ for $j=1,\ldots,C$ is the set of {\em control Hamiltonians} and the $C-$tuple of time functions $c=(c_1(t),\ldots,c_C(t))$ is the control vector.

A quantum system, described by a general density matrix $\rho$, defined on an $D-$dimensional Hilbert space ${\cal H}_N$ of $N$ particles and having a control set ${\cal C}$ is said to be density matrix controllable iff for every unitary transformation $U$ on ${\cal H}_N$ there exist at least one control vector $c$ and a time $t$ such that  
\begin{equation}
U=\exp\left(-\frac{i}{\hbar}
\sum_{j=1}^C\,\int_0^t\,\dd\tau c_j(\tau) H_j 
\right).
\end{equation}
Henceforth, Density Matrix Controllability will be referred to as \OSC; a system exhibiting \OSC will be said to be an \OSC system. It is well-known\cite{Dalessandro2007} that the system is \OSC if the control Lie algebra, which is the one ${\cal A_L(C)}$ generated by ${\cal C}$, spans the special unitary algebra ${\mathfrak su}(D)$ where $D$ is the dimension of the Hilbert space associated to the physical system. 

The available control Hamiltonians in the 2MQIHE are those corresponding to 1 qubit interactions with  external magnetic fields:
\begin{equation}H_{1\,{\rm through}\,6}= \sigma^{(1,2)}_{x,y,z},\end{equation}
and the Heisenberg interaction:
\begin{equation}
H_7=\sum_{j=x,y,z} \sigma^{(1)}_j\, \sigma^{(2)}_j.
\end{equation}
It can be shown\cite{Dalessandro2007} that the 2-qubit system endowed with the control set:\begin{eqnarray}\label{ctmqihe}
{\cal C}_2=\{&\sigma_x^{(1)}&\otimes \mathbbm{1}^{(2)}, \sigma_y^{(1)}\otimes \mathbbm{1}^{(2)},\sigma_z^{(1)}\otimes \mathbbm{1}^{(2)},\nonumber\\
&\mathbbm{1}^{(1)}&\otimes \sigma_x^{(2)},\mathbbm{1}^{(1)}\otimes \sigma_y^{(2)},\mathbbm{1}^{(1)}\otimes \sigma_z^{(2)},\nonumber\\
&\sigma_x^{(1)} & \otimes \sigma_x^{(2)}+\sigma_y^{(1)} \otimes \sigma_y^{(2)}+\sigma_z^{(1)} \otimes \sigma_z^{(2)}\},
\end{eqnarray}
is density matrix controllable. This implies that the system can be steered from any input state to any output state provided that their density matrices are unitarily equivalent. Two density matrices are unitarily equivalent iff their spectra share the same eigenvalues with equal multiplicity. Under these premises the system can be steered from one state to the other by a suitable choice of the control vector $c$. 

 Considering now the feedback version for the 2MQIHE, no further conditions need to be met. The control set ${\cal C}$ may steer the system from any post-measurement state to any other as long as both of them are unitarily equivalent. Note that the four possible outcomes determine four unitarily equivalent states which are equal up to at most two NOT-gates. Accordingly, controlability of feedback machines can be referred to that of swap engines.


\section{Controlled Thermalizability}
\label{sec:qthermo}
As has been established in the introduction to section \ref{sec:qcontrol}, it is necessary to tune the control Hamiltonians so that the state reached after the unitary evolution stage is in thermal equilibrium at temperature $T$. The central oval in Fig.\ref{fig:flow} represents such state. If the system is \OSC, as established in the previous section, the input and the thermal equilibrium states should be equal up to an unitary transformation. Now the problem can be stated as: given a control set ${\cal C}$ containing $C$ hermitian operators $H_1,\ldots,H_C$, is there a linear combination thereof whose thermal equilibrium state $\rho_\beta(H)$ is unitarily equivalent to any input state $\rho_i$?. Henceforth this property will be refered to as {\em controllable thermalizability} and abbreviated as \CT. In order to answer this question the following comments are in order:
\begin{enumerate}
\item The spectrum of $\rho_\beta(H)$ is the set ${\mathfrak S}=\{Z^{-1}\, e^{-\beta \lambda_\ell}
\}$ where $Z=\sum_\ell e^{-\beta \lambda_\ell}$ and $\lambda_\ell$ is an eigenvalue of $H$.
\item Considering that the traces of $\rho_i,\rho_\beta(H)$ are equal to 1, their spectra are equal iff those  of $-\beta^{-1}\log \rho_i$ and $H$ are shifted relative to each other by a real value which could be either positive, negative or zero. 
\end{enumerate}  
As a consequence, multiples of the identity can be appended to the control set ${\cal C}$  or added to any of its elements without changing neither \OSC nor \CT. Note that both concepts have been defined for systems with any dimensionality and control set ${\cal C}$.

Appendix \ref{sec:CTcase} shows that, in addition to being \OSC, 2MQIHE is \CT. This means that this engine can extract the maximum work $k_BT\,(\ln 2)\,\left[2-S(\rho_i)\right]$ from any input state $\rho_i$. 

The analysis of the \OSC and \CT problems presented in section \ref{sec:qcontrol} and appendix \ref{sec:CTcase} yield different results for $N>2$ and for $N=2$-qubit engines. With all 1-qubit Hamiltonians and a Heisenberg interaction Hamiltonian for each couple of qubits, the system is controllable, irrespective of the number $N$ of qubits involved. However, the result for the existence of a thermal equilibrium state unitarily equivalent to $\rho_i$ by a suitable combination of the control Hamiltonians is different. This can be inferred by noticing that for large $N$ the number of eigenvalues of $\rho_i$ scales as $2^N$, while the number of adjustable parameters for one-qubit Hamiltonians and two-qubit Heisenberg interactions scales as $N^2$. This consideration leads to the conclusion that no reversible process consisting of unitary and isothermal evolutions exists for systems of an arbitrary number of qubits when only the aforementioned control Hamiltonians are available. Moreover, even if not only one-qubit and Heisenberg interactions, but any two-qubit interactions are reckoned, the same conclusion holds, as is obvious because the scaling of the independent parameters also goes as $N^2$. Consequently, $k-$body interaction control Hamiltonians are needed. Notice the difference with the controllability problem, where only one and two-body interactions are sufficient for achieving \OSC, as can be easily inferred from the fact that universal sets of gates for $N$-qubit systems can be made of only one and two-qubit gates\cite{rmp}. On the other hand, in one-qubit systems, the $\sigma_z$  control Hamiltonian guarantees the \CT, yet failing to provide \OSC. Thus, \CT neither implies nor is implied by \OSC.

{ Notice that pure states can only become thermal states under Hamiltonians with infinite energy differences. For instance, in a 1MQIHE, a pure state is also a thermal state only if the magnetic field is infinite. In this paper we admit unrestricted linear combinations of control Hamiltonians and implicitly consider the limit of any succession of thermal states as a thermal state in its own right. This assumption allows pure states to be also thermal states.}
\section{USITIR Machines: Quantum Engines using Many-Body States}
\label{sec:qmany}
An $N-$qubit Hilbert space is equivalent to a $D=2^N$-dimensional qudit. In order to pursue our task of extending the previous study to many-body states we will next define a more abstract QIHE. Examining the one and two-qubit machines of section \ref{sec:qengine}, generalization to qudits and a wider set of engines can now proceed straightforward. We consider a physical device with the following elements:
\begin{enumerate} 
\item[(a)] an internal or system $D-$dimensional qudit $S$,
\item[(b)] a control set ${\cal C}$ containing $C$ hermitian operators $H_1,\ldots,H_C$ so that qudit ${S}$ can be made to evolve unitarily under a Hamiltonian  
\begin{equation}\label{laf}H=\sum_{j=1}^C c_j(t)\,H_j + f(t)\,\mathbbm{1}_{D\,\times\,D},\end{equation}
for any control vector $c=(c_1(t),\ldots,c_C(t))$ whose components are arbitrary functions of time. The elements of ${\cal C}$ play a dual role. They steer the density matrix of the system according to a definite strategy, while at the same time provide or store the energy interchanged as work in the process. In addition, $f(t)$ is any function of time, possibly unknown, that, as will be seen, plays no role in the final energy balance,
\item[(c)] a thermal reservoir $R(T)$ at temperature $T$,  
\item[(d)] a reversible isothermal mechanism whereby the Hamiltonian can be taken from any of the values attainable in item (b) to a multiple of the identity, with the extraction of work $W=-\Delta F$ where $F$ is the Helmholtz free energy of qudit ${S}$. This process leads to a completely depolarized final state for ${S}$, 
\item[(e)] a not-completely depolarized $D-$dimensional qudit $A$, referred to as {\em ancilla} or {\em input} qudit, with a mechanism to either be swapped with $S$ or hold the result of a generalized $CNOT(S,A)$ operation\cite{Bombin2005,Wilmott2012}, followed by a Von Neumann measurement of $A$ in the computational basis, as explained for feedback versions in previous sections.
\end{enumerate}
Besides, the engine must work in cycles, so that the initial and final states of qudit ${S}$ are completely depolarized, the initial and final Hamiltonians are the same and the final state of qudit ${A}$ is also completely depolarized. Finally, we suppose that the machine can also implement an irreversible thermalization whereby the system qudit $S$ with Hamiltonian $H$ in a state $\rho\neq\rho_\beta(H)$ is put in thermal contact with $R(T)$; in this situation we assume that the equilibrium state $\rho_\beta(H)$ is reached without performing any work and keeping the Hamiltonian constat in the process. 

\begin{figure}
\begin{center}
\includegraphics[scale=1]{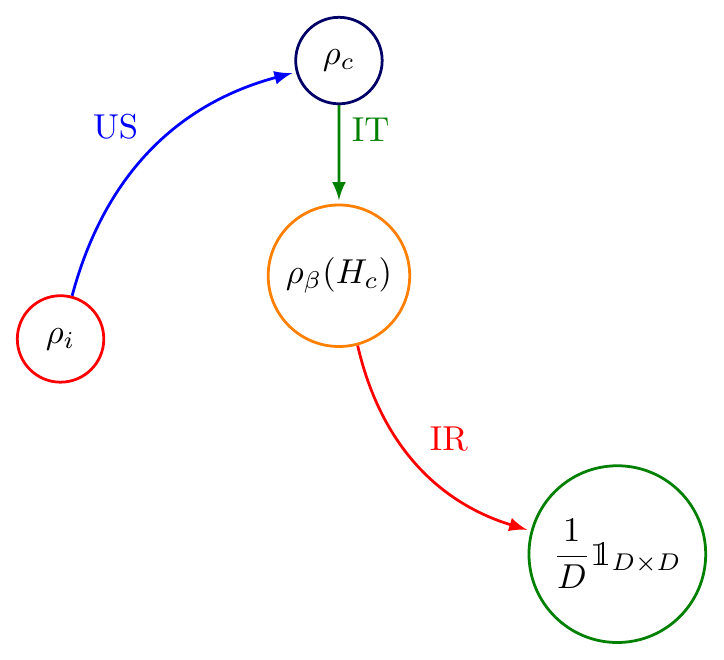}
\end{center}
\caption{(Color online) Schematic view of USITIR engines workflow. In the first stage (US) the input state $\rho_i$ is unitarily steered to a state $\rho_c$ with Hamiltonian $H_c$, from which an irreversible thermalization (IT) under constant Hamiltonian and with no work extraction takes the system to the equillibrium state for Hamiltonian $H_c$ given by $\rho_\beta(H_c)=\frac{e^{-\beta\,H_c}}{Z}$. Finally, the Hamiltonian is progressively turned off in an isothermal process (IR) leading to a completely depolarized state. }
\label{fig:USITIR-flow}
\end{figure}

After a swap or feedback operation, we assume a three-stage process, as depicted in Fig.\ref{fig:USITIR-flow}, defined by:
\begin{enumerate}
\item[(US)] Unitarily Steered evolution, driven by a suitable choice of the control vector $c$ that takes qudit ${S}$ from the initial state $\rho_i$ to another one $\rho_c$, chosen to maximize the extractable energy, as will be analyzed in section \ref{sec:qwork}. If the control set allows it, $\rho_c$ will be a thermal equilibrium state for one of the available Hamiltonians.
\item[(IT)] Irreversible Thermalization, where the state is taken to thermal equillibrium without any work extraction and keeping the Hamiltonian constant. This stage is irrelevant if $\rho_c$ is a thermal equilibrium state.
\item[(IR)] Isothermal Relaxation, that takes the system reversibly and isothermally to the completely depolarized state under the initial Hamiltonian. 
\end{enumerate}
An Information Heat Engine fitting into this description will be called an {\em USITIR} machine. 

Given premises (a) through (e), the three USITIR stages and their sequential order are unavoidable. Once the thermal equilibrium is reached, no work gain can be achieved by alternating unitary and isothermal steps. Unitary evolution must take the first turn and irreversible thermalization inevitably proceeds in between.

Qudits systems are introduced with the intention to naturally accommodate $N$-particles, particulary $N$-qubit systems with quantum statistics. One-qubit, two-qubit or $N-$qubit MQIHEs of distinguishible particles fit into $D=2,D=4,D=2^N$-dimensional qudit USITIR machines, respectively. Moreover, $N$-bosonic qubits are naturally modeled with $D=(N+1)$-dimensional qudits, whereas a 2 fermion qubit system  is equivalent to a trivial one-dimensional qudit. 


In statistical physics particle interactions are often neglected and all particles are assumed to interact with external Hamiltonians. For one particle systems where the Hilbert space is $d$-dimensional, we define the control set ${\cal L}_1$ as the one formed by all $d^2-1$ generators of $SU(d)$. Assuming $N$ distinguishable particles with a $d$-dimensional Hilbert space for each, we define the {\em local independent} control set ${\cal L}_N$ as the union of all $N$ control sets ${\cal L}^{(k)}_1$  for $k=1,\ldots N$, each one acting on its part of the product space. We will pay special attention to the case of $N=2$ qubits, with $d=2$, so that 
\begin{eqnarray}
{\cal L}_2:=&\{&\sigma_x^{(1)}\otimes \mathbbm{1}^{(2)}, \sigma_y^{(1)}\otimes \mathbbm{1}^{(2)},\sigma_z^{(1)}\otimes \mathbbm{1}^{(2)},\nonumber\\
&&\mathbbm{1}^{(1)}\otimes \sigma_x^{(2)},\mathbbm{1}^{(1)}\otimes \sigma_y^{(2)},\mathbbm{1}^{(1)}\otimes \sigma_z^{(2)}\,\,\,\}.
\end{eqnarray}

Hilbert spaces for bosonic or fermionic systems can be envisioned as subspaces of the distinguishable particles ones. The pure states that span the Hilbert subspace feature symmetry or antisymmetry with respect to two-particle swapping. On the other hand, Hamiltonians for identical particle systems without mutual interactions are always symmetric under the same operation. This implies that a bosonic or fermionic density matrix always remains bosonic or fermionic, respectively, under any unitary transformation built from the available control set. Consequently, for $N$ identical bosonic or fermionic USITIR engines it may  prove practical to describe the control set or the density matrices using a basis for the larger system, i.e., the one which corresponds to the same number of distinguishable particles. 

As all identical particles are assumed to be indistinguishable, they experience the same Hamiltonians. We will be specially interested in non-interacting situations and define a special control set ${\cal G}_N$ for systems of $N$ identical particles, each having an individual $d$-dimensional Hilbert space. Working in the $N-$th tensor power of the one-particle Hilbert space,  ${\cal G}_N$ is the {\em local common} control set of $d^2-1$ elements, each one being the sum $\sum_{k=1}^N\,\gamma_j^{(k)}$, where $\gamma_j^{(k)}$ is the $j-$th generator of $SU(d_1)$ acting on the $k$-factor space. For 2-qubit systems, we have
\begin{eqnarray}\label{def:gdos}
{\cal G}_2:=&\{&\sigma_x^{(1)}\otimes \mathbbm{1}^{(2)}+\mathbbm{1}^{(1)}\otimes \sigma_x^{(2)},\nonumber\\ && \sigma_y^{(1)}\otimes \mathbbm{1}^{(2)}+\mathbbm{1}^{(1)}\otimes \sigma_y^{(2)},\nonumber\\
&& \sigma_z^{(1)}\otimes \mathbbm{1}^{(2)}+\mathbbm{1}^{(1)}\otimes \sigma_z^{(2)}\,\,\,\}.
\end{eqnarray}

A final type of control set for $N$-qubit systems is given by the singleton
\begin{equation}\label{def:FN}
{\cal F}_N:=\{\sum_{k=1}^N\sigma_z^{(k)}\},
\end{equation}
being $\sigma_z^{(k)}$ the operator that performs a $\sigma_z$ in the $k$-th factor subspace.  This control set will prove useful when considering Szilard engines, as is explained in section \ref{sec:qwork}. We will use the control sets ${\cal G}_N,{\cal F}_N$ in USITIR engines for both distinguishable and identical particles.

In the particular case where \OSC and \CT are fulfilled, that is, the control set ${\cal C}$ allows the engine to be steered to a thermal equilibrium state in the US stage, the machine is able to extract the maximum work compatible with Thermodynamics and is said to be {\em optimal}; it will obtain a work given by the difference between the energy decrease in the US process and the free energy increase in the IR stage. Thus, the work obtained, bearing in mind that the increment of the energy in a full cycle must be null, reads:
\begin{equation}\label{optim}
W=k_B\,T\,(\ln 2)\,(\log_2 D -S(\rho_i)),
\end{equation}
where $\rho_i$ is the initial state of ${A}$. According to the previous discussion, 1MQIHE, 2MQIHE and, more generally, all \OSC and \CT machines are optimal. A general expression for the extractable work in non-optimal USITIR engines is given in the next section.

\section{Extractable Work and Uncontrollable Entropies}
\label{sec:qwork}

As has been previously shown, the availability of control Hamiltonians that allow to accomplish both \OSC and \CT guarantees a possibility for the extraction of an amount of work given by eq.\eqref{optim}. However, this may not always be the case. It could happen that the control Hamiltonians fail to provide a reversible way to a thermal state for input density matrix $\rho_i$. We assume then that at some point of the process there must be an irreversible thermalization step with neither work extraction nor change of the Hamiltonian.  This irreversible stage contributes a work penalty $W_p$ given by the work performed by a reversible process between the same initial and final states and Hamiltonians, which is\cite{sagawa_ueda09}:
\begin{equation}
W_p = k_B T \,(\ln 2)\, S(\rho_1\,||\,\rho_2),
\end{equation}
where $\rho_1,\rho_2$ are the density matrices right before and immediately after the irreversible thermalization step, respectively. We may let $\rho_1$ take any value reachable from the input density matrix $\rho_i$ with the available control set ${\cal C}$ and $\rho_2$ vary along all thermal states defined with such control Hamiltonians. Let us then define the {\em uncontrollable entropy} of $\rho_i$ under ${\cal C}$ by:
\begin{equation}\label{def:Su}
S_u(\rho_i,{\cal C}):=\mathop{\rm min}_{\rho_1,\rho_2} S(\rho_1\,||\,\rho_2),
\end{equation}
where minimization is taken over the previously defined ranges for $\rho_1,\rho_2$. This magnitude, always non-negative, is null for the 1MQIHE and 2MQIHE described in subsection \ref{subsec:two-qengine}, on account of its \OSC (see section \ref{sec:qcontrol}) and \CT (see section \ref{sec:qthermo}). 


According to definition \eqref{def:Su}, the maximum extractable work obtainable from an USITIR engine is:
\begin{equation}\label{eq:exw}
W(\rho_i,{\cal C}) = k_B T\,(\ln 2)\,(\log_2 D -S(\rho_i)-S_u(\rho_i,{\cal C})).
\end{equation}  

Next we analyze a control set ${\cal C}$ acting on a two-qubit system in two particular cases: a) ${\cal C}={\cal L}_2$ and b) ${\cal C}={\cal G}_2$. 

In case a) it is evident that the range for $\rho_2$ is the set of factorizable states and thus the uncontrollable entropy is the sum of the entropies of the reduced states minus that of the initial state (see Appendix \ref{sec:local} for extended calculation)  

\begin{equation}
S_u(\rho_i,{\cal L})=S(\rho_i^{(1)})+ S(\rho_i^{(2)})- S(\rho_i),
\end{equation}
where $\rho_i^{(1)},\,\rho_i^{(2)}$ are the reduced states of qubits $A_1,A_2$ in the initial bipartite state $\rho_i$.

For case b) $\rho_2$ ranges over all factorizable states with equal factors and thus the uncontrollable entropy is the sum of the entropies of the averaged reduced states minus that of the initial state (see Appendix \ref{sec:local} for extended calculation)  

\begin{equation}\label{ce-global}
S_u(\rho_i,{\cal G})=2 \, S(\frac{\rho_i^{(1)}+\rho_i^{(2)}}{2})- S(\rho_i).
\end{equation}

In feedback  USITIR engines the situation is somewhat different, since the problems of controlability and thermalizability appear for each of the equally likely $P$ possible outcomes $\rho_{i,\,p}\,p=1,\ldots P$ of the measurement, where $P=D$ if the state of qudit $S$ before performing the $CNOT(S,A)$ operation is completely depolarized. The analysis leads to several problems each of them being a particular case of the swap machine situation. The extractable work results in an average of the results for the $P$ possible outcomes, given by
\begin{equation}\label{w-fb}
W_{\sc fb}(\rho_i,{\cal C}) = k_B T\,(\ln 2)\,(\log_2 D-S(\rho_i)-\frac 1 P \sum_{p=1}^PS_u(\rho_{i,\,p},{\cal C})).
\end{equation}


Using  eq.\ref{us-elen} of appendix \ref{sec:local} for the uncontrollable entropy, one can conclude that when many-qubit machines only have local independent control Hamiltonians, i.e. the control set is ${\cal L}_N$, then they are optimal only if 
\begin{equation}\label{suman}
S(\rho_i)=\sum_{k=1}^N S(\rho_i^{(k)}),
\end{equation}
that is fulfilled only for a particular class of input sates, namely those factorizable in their $N$ partial states. The same result is shared by swap and feedback machines, since in the latter the factorizability is not affected by the possible $NOT$ gates that link the measurement outcomes.

If, besides being single qubit, the control Hamiltonians are common to all qubits, i.e. control set is ${\cal G}_N$, then according to the uncontrollable entropy given by eq.\eqref{us-gn}, the engine is  optimal only if:
\begin{equation}
S(\rho_i)=\,N\,S\left(\frac{1}{N}\sum_{k=1}^N{\rho_i^{(k)}}\right),
\end{equation}
which, on account of the concavity of Von Neuman entropies, is satisfied only if $\rho_i$ represents a factorizable state of equal components. For feedback machines this is impossible, because the internal qubits start from a completely depolarized state and, consequently, all post-measurement states are equally likely.

The previous definitions of extractable work and uncontrollable entropy also apply to $N$-particle systems with quantum statistics, provided they are treated within the correct Hilbert space. Systems of $N$ fermionic qubits are not very interesting: for $N=1$ statistics play no role, for $N=2$ the Hilbert space is trivially unidimensional and no systems exist for $N>2$. Accordingly, further attention will be paid to the $N$-qubit bosonic cases only, which are naturally accommodated in $D=(N+1)$-dimensional qudit USITIR engines. In addition, a ${\cal F}_N$ control set will be assumed. The relevant action of the US stage reduces to driving the energy difference of the two levels from an initial value to another one from where the IT process is launched. A derivation of the extractable work for $N$-bosonic qubits is given in Appendix \ref{sec:bosons}.  


So far, we have considered that the quantum many-body states in the system of the USITIR engine are particle-distinguishable. Now, we are going to  derive the corresponding formulas of the optimal extractable work for identical particles in SZE engines and compare with recent results \cite{KSDU11, PDGV12}.
{ Indeed, SZEs are the best-known Information Heat Engines. A one-particle SZE is basically a hollow cylinder whose inner space is divided into two parts by a barrier $B$. The motion of $B$ is externally controlled according to the outcomes of a measurement on the positions of the particles. The displacement of the barrier is an unitary process which starts at the center of the cylinder and ends at a suitably chosen position. Then, after the particles are allowed to redistribute, the barrier is restored to its initial position and a new cycle begins. Focusing on the unitary displacement of $B$, the temperature is presumed  to be low enough to assume that the particle is in a superposition of the fundamental left and right states $\ket{L},\,\ket{R}$, or a density matrix built on the  space spanned by them. Thus, the system is reduced to a qubit for one particle or, more generally, to $N$ qubits if there are $N$ particles in the cylinder. As the barrier moves, the energies of $\ket{L},\,\ket{R}$ change, being the Hamiltonian of the system  a linear combination of the identity and $\sigma_z$ operators. Moreover, the position of $B$ is the only parameter that determines the Hamiltonian of the system.  Therefore, this stage is equivalent to the US step of an N-qubit, ${\cal F}_N$ control set, USITIR engine. The last stage is an isothermal reversible restoration of the barrier to its initial stand, that matches exactly the IR step of the N-qubit, ${\cal F}_N$ USITIR machine. The transition from the first to the last stage may always proceed reversibly for the $N=1$-particle case. As for USITIR engines, every cycle yields an optimal $k_B T \,\ln 2$ work output.  For more than one particle, if no further steering resources are available, the uncontrollable entropies computed in appendixes \ref{sec:FN} and \ref{sec:bosons} show that there must be another stage, called IT in the USITIR set-up, between US and IR, that contributes a work penalty that is also obtained in appendixes \ref{sec:FN} and \ref{sec:bosons}. 

In order to draw some results for particular examples, we next consider the $N=2$-qubit  feedback engine with the ${\cal F}_2$ control set, under distinguishable, bosonic and fermionic statistics.}  

In feedback engines, if the particles are distinguishable, eq.\eqref{w-fb} for the obtainable work reads:
\begin{equation}\label{ws}
W_{\sc Fb}(\rho_i,{\cal F}_2) = k_B T\,(\ln 2)\,(2-\frac 1 4 \sum_{p=1}^4S_u(\rho_{i,\,p},{\cal F}_2)).
\end{equation}
In order to compute the uncontrollable entropy, we use the general expression given by eq.\eqref{unod} obtained in appendix \ref{sec:FN}
\begin{equation}\label{si}
S_u(\rho_{i,\,p},{\cal F}_2)=2 \,S(\frac{\rho_{i,d,p}^{(1)}+\rho_{i,d,p}^{(2)}}{2}),
\end{equation} 
where $\rho_{i,d,p}^{(k)}$ is the $k$-th reduced matrix corresponding to the $p-$th measurement outcome, decohered\footnote{the density matrix $\rho$  {\em decohered} in the computational basis refers to the state $\rho_d:=\sum_{k=1}^D\ket{k}\bra{k}\,\rho\,\ket{k}\bra{k}$, that is, the same $\rho$ with all non-diagonal terms removed. } in the computational basis. Considering pure state ancillas, which implies that after measuring qudit $A$, qudits $A,S$ are in the same pure state, eq.\eqref{si} evaluates to $0$ when $\rho_{i,p}$ represents post-measurement states in which the two particles are on the same side and to $2$ when they are on different sides. Substituting into eq.\eqref{ws}, we obtain
\begin{equation}\label{wsz}
W_{\sc Fb}(\rho_i,{\cal F}_2) = k_B T\,(\ln 2),
\end{equation}
in agreement with Ref.\cite{KSDU11}. 

The fermionic case is trivial and the bosonic one can be easily discussed taking into account the results of appendix \ref{sec:bosons}. According to eq.\eqref{w-fb} and reading from Table \ref{tab:bos}, the maximum extractable work in the bosonic 2-qubit system, when it is fed with pure state ancillas, is:
\begin{equation}\label{wszd}
W_{\sc Fb}(\rho_i,{\cal F}_2) = \frac 2 3 \,k_B T\,(\ln 3),
\end{equation}
again in agreement with Ref.\cite{KSDU11}. If more control Hamiltonians were available, the uncontrollable entropy could be made null and more work could be extracted. Assuming $\OSC$ and $\CT$, the quantum Szilard engine would yield the optimum value given by eq.\eqref{optim}, which, if ancillas are fed in a pure state, reads
\begin{equation}\label{wszk}
W_{\sc Fb}(\rho_i,{\cal F}_2) = k_B T\,\ln D,
\end{equation}
where $D=4,3,1$ for distinguishable, bosonic and fermionic two-qubit systems, respectively, in agreement with Ref.\cite{PDGV12}.

\section{Conclusions}
\label{sec:qconclusions}
This paper has presented a concept model for a Magnetic Quantum Information Heat Engine. As such, it extracts heat from a thermal reservoir and converts it into electrical work  at the expense of increasing the entropy of input qubits. In addition, it features some interesting properties:
\begin{enumerate}
\item It is scalable in the sense that it deals with information contained in any number of arbitrarily entangled qubits. For two-qubit systems a physical model has been proposed that saturates the thermodynamic limit. For a higher number of qubits it has been proved that many-body interactions are needed.
\item The workflow for the swap model does not include any measurement or feedback system, unlike most modern descriptions of IHE. Nevertheless, a feedback version has been described to show that its study refers to the one without feedback.   
\item A generalization, called USITIR engine, has been described that includes other well-known models.
\end{enumerate}

Besides, we have described a general framework for analyzing the performance that can be expected from a wide class of Quantum Information Heat Engines. It focuses on the set of control Hamiltonians that must be studied in two particular problems. The first one is assuring its \OSC. The second one is more original and has been named {\em controllable thermalizability}. It refers to the search of  a Hamiltonian, within the set of allowable ones, that defines a thermal equilibrium state at temperature $T$ which is unitarily equivalent to the input state of the ancilla. It turns out that neither of the properties, \CT and \OSC, implies the other, as shown in section \ref{sec:qmany}. 

 For non-optimal machines, a measure of the penalty in the extractable work has been formulated as a function of a newly defined quantity: {\em uncontrolable entropy}.  The generalization presented in section \ref{sec:qmany} predicts the limitation of engines with only one-qubit independent Hamiltonians or one-qubit common ones. Different kinds of quantum statistics effects have been shown to fit nicely into the general model. Detailed calculations developed in appendixes \ref{sec:local}, \ref{sec:FN} and \ref{sec:bosons}  quantify their maximum extractable work and identifies the cause of their limitation for $N-$qubit engines under several control sets and for different quantum statistics. The conclusion, as formulated at the end of section \ref{sec:qwork}, is that neither independent nor common local Hamiltonians allow optimal work extraction from arbitrary $N$-qubit inputs. Only for particular classes of inputs, as described in section \ref{sec:qwork}, swap and feedback engines can be optimal. However, when control Hamiltonians are common, irrespective of the input states, no feedback engine can be optimal. This type of engine represents the case of many magnetic qubits in a common induction field or many particles in a cylinder, separated by the same barrier.


\begin{acknowledgments}

We thank the Spanish MICINN grant FIS2009-10061, FIS2012-33152,
CAM research consortium QUITEMAD S2009-ESP-1594, European Commission
PICC: FP7 2007-2013, Grant No.~249958, UCM-BS grant GICC-910758.

\end{acknowledgments}

\appendix
\section{Work extraction calculation for 1MQIHE}
\label{sec:energ}

In this appendix we proceed to obtain an expression for the electric power delivered to a battery in the different stages of a 1MQIHE. Then we use it to draw a final balance of energy in a full cycle. 

We start from the reciprocity theorem\cite{Insko1998}, as is often done in Nuclear Magnetic Resonance literature. Let $\vec B_1=B_1 \vec u_z$ be the magnetic induction generated at the place of the spin atom, assuming that the qubit $S$ is an atomic spin, by a unit current running through $C$, so that the magnetic field $\vec B$ at $S$ reads $\vec B=I_C B_1\vec u_z$. Then the electromotive force ${\cal E}'$ induced on $C$ by the magnetic moment $\vec{\mu}$ of the atom is:
\bel{reciproc}
{\cal E}' = -\der{(\vec{B}_1\cdot\vec \mu)}{t} = -B_1\der{\mu_z}{t},
\eel 
and the total electromotive force for circuit $C$ is 
\begin{equation}{\cal E}=-L\der{I_C}{t} + {\cal E}'.\end{equation}
Thus, the electric power supplied to the battery now reads
\begin{equation}
P_s=I_C {\cal E} = -L I_C \der{I_C}{t}-I_C B_1\der{\mu_z}{t}
\end{equation}  
and taking into account that $B=I_C B_1$, 
\bel{Ps}
P_s=-\der{R}{t} + \mu_z \der{B}{t}\,\mbox{with }R= \frac 1 2 L I_C^2 +  \mu_z B.
\eel

According to eq.\eqref{Ps}, the electric battery connected to the source keeps an energy in excess of its initial value given by:  
\begin{equation}
E_s=\int_0^t \dd\tau\,P_s(\tau)=-\Delta R\,+\int_0^B\,\dd B\,\mu_z(B), 
\end{equation}
and in a cycle:
\begin{equation}
E_s=\oint\,\dd B\,\mu_z(B).
\end{equation}

\begin{figure}[t]
\includegraphics[width=0.27\textwidth]{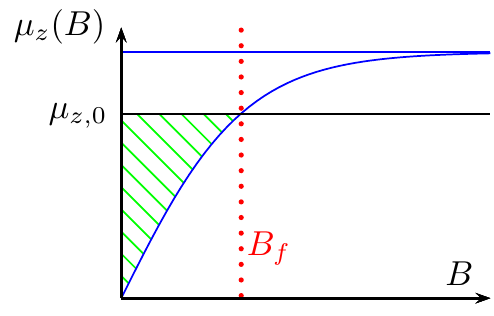}
\caption{(Color online) Brillouin magnetization curve for qubit $S$. The cycle of the 1MQIHE corresponds to a clockwise loop around the green hatched zone whose area represents the extracted work.}
\label{fig::energia}
\end{figure}

Fig. \ref{fig::energia} shows the energy gain, which corresponds to the hatched area, when the ancilla is fed in a state $\frac 1 2 ({\mathbbm 1}+c\sigma_z)$, so that, after swapping, qubit $S$ has a magnetic moment $\mu_{z,0}=c \mu_M $. Now we apply these expressions to the three stages previously described. 
\begin{itemize}
\item[i) ] At this stage, the $I_C,B$ are constant and so $P_s$ is zero. There is no power flowing from or into the battery and the function $R$ is also zero.
\item[ii) ] Now the $R$ function increases, implying a negative energy flow into the battery. There is also a positive contribution $\mu_z \dd B$ which does not compensate for the negative one. The energy of the battery is incremented by 
$
\Delta E_{2} = -\Delta R + \mu_{z,0} B_{f}.
$
\item[iii) ] At this stage the $R$ component falls back to zero, restoring the negative contribution of the previous stage. There is also a negative term coming from the negative value of $\dd B$. As there is thermal equilibrium, $\mu_{z}$ depends on $B$. The state of qubit $S$ is:
\begin{equation}
\rho_\beta(-\mu_M\,B\,\sigma_z) =  \fra{e^{\beta \mu_M\, B } \ketup \braup \,+\,e^{ -\beta \mu_M\, B }\,\ketdown\bradown}{e^{-\beta \mu_M\, B }+e^{\beta \mu_M\, B }},
\end{equation} 
so that $
\mu_{z} = \mu_M \tanh \beta \mu_M B
$
and the energy stored in the battery is incremented by $
\Delta E_{3} = \Delta R - \frac 1 \beta \ln \cosh \beta\mu_M B_f.
$
\end{itemize}
Adding all contributions we have:
\begin{equation}
\Delta E_s = \mu_{z,0} B_{f}- \frac 1 \beta \ln \cosh \beta\mu_M B_f.
\end{equation}
In order to find the optimum value for $B_f$ we derive the expression and immediately find that $
\mu_z= \mu_M \tanh \beta \mu_M B_f
$ 
, so that the value of $B_f$ is the one that makes the $S$ qubit state $\rho_{S,1}$ to be in equilibrium with the reservoir at temperature $T$. This result could have been anticipated on account that it is the only one that makes the process reversible. The resulting energy is:
\bel{fee}
\Delta E_s = \mu_M B_{f}\,\tanh \left( \beta \mu_M B_f\right) - \frac 1 \beta \ln \cosh \left(\beta\mu_M B_f\right).
\eel
Now we will explore the relation of \eqref{fee} with the initial entropy of the ancilla, which is the same as that of $\rho_{S,1}$. Using the fact that $B_f$ makes $\rho_{S,1}$ in thermal equilibrium with a reservoir at temperature $T$, we can express the entropy $S(\rho_{S,1})$ as a function of $B_f$ 
\begin{eqnarray}
S(\rho_{S,1})=&&\frac{e^{\beta\mu_MB_f}}{Z\,{\ln 2}}\left(-{\beta\mu_MB_f} +\ln Z\right)\nonumber\\
&&+\frac{e^{-\beta\mu_MB_f}}{Z\,{\ln 2}}\left({\beta\mu_M \, B_f} +\ln Z\right),
\end{eqnarray}
where $Z=2\cosh\beta\mu_M B_f$. Further calculations yield:
\bel{entropia}
S(\rho_{S,1})=\frac{\ln \left(2\cosh \beta\mu_M B_f\right)-\beta\mu_M B_f\,\tanh \beta\mu_M B_f}{\ln 2}.
\eel 
Combining \eqref{fee} and \eqref{entropia} we arrive at:
\bel{es}
\Delta E_s = \frac{\ln 2}{\beta}  \left[1-S(\rho_{S,1})\right],
\eel
which is equal to $k_B T \ln 2$ times the entropy decrease of the reservoir, assuming that it is measured in bits.

\section{Controllable Thermalizability for 2MQIHE}
\label{sec:CTcase}
In this appendix we show that the  control set ${\cal C}_2$, defined in eq.\eqref{ctmqihe}, acting on two-qubit engines, features \CT, in addition to \OSC. Considering that the spectrum of $\rho_\beta(H)$ is invariant under any transformation belonging to $SU(2)^{(1)}\otimes\,SU(2)^{(2)}$ and that any linear combination of the first(second) three operators of ${\cal C}_2$ can be reduced to a multiple of $H_3=\sigma_z^{(1)}\otimes \mathbbm{1}^{(2)}$ ($H_6=\mathbbm{1}^{(1)}\otimes \sigma_z^{(2)}$) by some such transformation, we can limit our search to linear  combinations of $H_3,H_6,H_7$ that match the spectrum of the initial density matrix $\rho_i$ up to an additive constant. This is equivalent to finding a linear combination of $H_3,H_6,H_7,\mathbbm{1}_{4\times 4}$ which is unitarily equivalent to $\rho_i$.

It proves useful to write the Hamiltonian as  
\begin{equation}\label{intrin}
H=c_1 \frac{\mathbbm{1}_{4 \times 4} - H_3}{2} + c_2 \frac{\mathbbm{1}_{4 \times 4} - H_6}{2} +c_3 \frac{\mathbbm{1}_{4 \times 4} - H_7}{2} + c_4 \mathbbm{1}_{4 \times 4} 
\end{equation}
and solve for $c_1,c_2,c_3,c_4$ under the condition that $H$ and $\rho_i$ share the same spectrum. In matrix form, the Hamiltonian reads 
\begin{eqnarray}
H&=&c_1\left(
\begin{array}{cccc}
0&0&0&0\\
0&0&0&0\\
0&0&1&0\\
0&0&0&1
\end{array}
\right)
+
c_2\left(
\begin{array}{cccc}
0&0&0&0\\
0&1&0&0\\
0&0&0&0\\
0&0&0&1
\end{array}
\right)\nonumber\\
& & 
+c_3 \left(
\begin{array}{cccc}
0&0&0&0\\
0&1&-1&0\\
0&-1&1&0\\
0&0&0&0
\end{array}
\right)
+c_4\left(
\begin{array}{cccc}
1&0&0&0\\
0&1&0&0\\
0&0&1&0\\
0&0&0&1
\end{array}
\right),
\end{eqnarray}
whose eigenvalues are
\begin{equation}\begin{array}{c}
c_4,c_4+c_1+c_2,\\
c_4+\displaystyle\frac{c_1+c_2+2c_3}{2} \,\pm\,\displaystyle\frac{\displaystyle\sqrt{(c_1+c_2-2c_3)^2 - 4 c_1\,c_2}}{2},
\end{array}
\end{equation}
henceforth referred to as $\lambda_1(c),\lambda_2(c),\lambda_3(c),\lambda_4(c)$, respectively, being $c=(c_1,c_2,c_3,c_4)$. They should match the eigenvalues of the density matrix $\rho_i$, which will be denoted by $\eigrho_1,\eigrho_2,\eigrho_3,\eigrho_4$, ordered as $\eigrho_1\leq\eigrho_2\leq\eigrho_3\leq\eigrho_4$ or any permutation thereof. Next we show that  there always exists at least one real solution for $c_1,c_2,c_3,c_4$  in the system of equations defined by 
\begin{equation}
\left\{\begin{array}{rcl} 
\lambda_1(c)&=&\eigrho_2\\ 
\lambda_2(c)&=&\eigrho_3\\
\lambda_3(c)&=&\eigrho_1\\
\lambda_4(c)&=&\eigrho_4 .
\end{array}\right.
\end{equation}
This can be inferred by considering the solution $c_4=\eigrho_2,\,c_3=\frac{\eigrho_1+\eigrho_4-\eigrho_2-\eigrho_3}{2}$ and $c_1,c_2$ determined by the system:
\begin{equation}\label{sistema}
\left\{\begin{array}{rcl}
c_1+c_2&=&\eigrho_3\\
c_1-c_2&=&\pm\sqrt{(c_4-c_1)^2-\left({\eigrho_1+\eigrho_4-\eigrho_2-\eigrho_3}\right)^2}.
\end{array}
\right.
\end{equation}
Eq,\eqref{sistema} defines a real solution for $c_1,c_2$ iff 
\begin{equation}
(\eigrho_1-\eigrho_4)^2 \geq (\eigrho_1+\eigrho_4 - \eigrho_2-\eigrho_3)^2,
\end{equation}
that, bringing into account the order $\eigrho_1\leq \eigrho_2\leq \eigrho_3\leq \eigrho_4$, is always fulfilled, proving the controllable thermalizability of the swap version of 2MQIHE.

It is straightforward that the previous result extends also to the feedback version of 2MQIHE, on account of its \OSC and realizing that all four post-measurement states are unitarily equivalent.

\section{Uncontrollable entropies for systems with local control Hamiltonians}
\label{sec:local}

In this appendix we present a derivation of the uncontrollable entropies for 2-qubit systems under the control sets ${\cal L}_2,{\cal G}_2$, as defined in section \ref{sec:qwork}. A generalization to $N$-qubit systems will follow immediately. 

First we study the case ${\cal C}={\cal L}_2$ in an USITIR engine. At the irreversible thermalization stage no work is assumed to be interchanged. Thus, work is only accounted for in the first and the last stages. For the first one we can write:
\begin{equation}\label{work-one}
W_1=\Traza{\rho_i\,H_i - \rho_c\,H_c},
\end{equation} 
where $\rho_c,H_c$ are the density matrix and the Hamiltonian at the end of the first stage. Recalling that the Hamiltonians allowed by ${\cal L}_2$ can be decomposed as 
\begin{equation}\label{factores}
H=H^{(1)}\otimes \mathbbm{1}^{(2)} + \mathbbm{1}^{(1)} \otimes H^{(2)},
\end{equation} 
we can rewrite eq.\eqref{work-one} as
\begin{equation}\label{work-two}
W_1=\Traza{\rho_i\,H_i - \rho_c^{(1)}\,H_c^{(1)}- \rho_c^{(2)}\,H_c^{(2)}}.
\end{equation} 
The work extracted at the last stage, taking into account that it is an isothermal process and decomposition eq.\eqref{factores}, is:
\begin{eqnarray}\label{work-three}
W_2= &&\Traza{ \rho_\beta(H_c^{(1)})\,H_c^{(1)}+ \rho_\beta(H_c^{(2)})\,H_c^{(2)} - \rho_i\,H_i} \nonumber \\ && + k_B T \,(\ln 2)\,\left[S(\rho_f) - S(\rho_d)\right],
\end{eqnarray}
where $\rho_d=\rho_\beta(H_c^{(1)})\otimes \rho_\beta(H_c^{(2)}),\rho_f=\frac 1 4\,\mathbbm{1}^{(1)}\otimes \mathbbm{1}^{(2)}$ are the density matrices at the beginning and the end of the last stage, respectively. 

Adding eq.\eqref{work-two} and eq.\eqref{work-three} we obtain the total work:
\hspace{-10mm}\begin{eqnarray}\label{work-four}\hspace{-10mm}
W&&=\Traza{ \left[\rho_\beta(H_c^{(1)})-\rho_c^{(1)}\right]H_c^{(1)}+ \left[\rho_\beta(H_c^{(2)})-\rho_c^{(2)}\right]H_c^{(2)} } +\nonumber \\ && k_B T\left[S(\rho_f^{(1)})+S(\rho_f^{(2)}) - S(\rho_\beta(H_c^{(1)}) - S(\rho_\beta(H_c^{(2)}))\right]\ln 2 . \nonumber\\
&& 
\end{eqnarray}
Notice the symmetry for the first and second qubits. Next we elaborate further on the part which refers to the first qubit
\begin{eqnarray}\label{work-four-o}
W^{(1)}&&=\Traza{ \left[\rho_\beta(H_c^{(1)})-\rho_c^{(1)}\right]\,H_c^{(1)}}\nonumber\\
&&+k_B T\left[S(\rho_f^{(1)}) - S(\rho_\beta(H_c^{(1)})\right]\ln 2 ,
\end{eqnarray}
where, considering that $H_c^{(1)}=-k_B T  (\log_2 Z_1 \mathbbm{1}^{(1)}+\log_2 \rho_\beta(H_c^{(1)})\ln 2$, we arrive at:
\begin{eqnarray}\label{work-four-od}
\frac{W^{(1)}}{k_B T \ln 2}&&=\Traza{ \left[-\rho_\beta(H_c^{(1)})+\rho_c^{(1)}\right]\,\log_2 \rho_\beta(H_c^{(1)})}\nonumber\\
&&-\Traza{\left[\rho_f^{(1)}\log_2(\rho_f^{(1)}) - \rho_\beta(H_c^{(1)})\log_2 \rho_\beta(H_c^{(1)})\right]} , \nonumber\\ &&
\end{eqnarray}
that can be written as:
\begin{equation}\label{work-four-ot}
\frac{W^{(1)}}{k_B T \ln 2}=S(\rho_f^{(1)})-S(\rho_i^{(1)}) - S(\rho_c^{(1)}||\rho_\beta(H_c^{(1)})).
\end{equation}
Maximization of $W^{(1)}+W^{(2)}$ can proceed independently for each term. Regarding that ${\cal L}_2$ allows complete controllability over each of the qubits independently and that relative entropies are non-negative, maximum work is:
\begin{equation}\label{work-ele}
W(\rho_i,{\cal L}_2)=k_B T \,(\ln 2)\,\left[2-S(\rho_i^{(1)})-S(\rho_i^{(2)})\right],
\end{equation}
that determines an uncontrollable entropy:
\begin{equation}\label{us-ele}
S_u(\rho_i,{\cal L}_2)=S(\rho_i^{(1)})+S(\rho_i^{(2)})-S(\rho_i).
\end{equation}
When the control set is ${\cal G}_2$, as defined in eq.\eqref{def:gdos}, all the previous derivation holds until eq.\eqref{work-four-ot}. Now maximization  must be made for the whole work $W=W^{(1)}+W^{(2)}$. With this goal we write the expression for it:
\begin{eqnarray}\label{workg}
\frac{W}{k_B T \ln 2}&&=  S(\rho_f)-S(\rho_i^{(1)})-S(\rho_i^{(2)}) \nonumber\\
&&  - S(\rho_c^{(1)}||\rho_\beta(H_c))- S(\rho_c^{(2)}||\rho_\beta(H_c)) ,
\end{eqnarray}
where any superindex for a magnitude which is equal for both subsystems has been suppressed. After expansion of the expressions for the relative entropies and simplification we arrive at:
\begin{equation}\label{workge}
\frac{W}{k_B T \ln 2}=  S(\rho_f)  + \Traza{\left[\rho_c^{(1)}+\rho_c^{(2)}\right]\log_2\rho_\beta(H_c))},
\end{equation}
which can be rewritten as
\begin{equation}\label{workgeg}
\frac{W}{2 k_B T \ln 2}=   S(\rho_f^{(1)})- S(\rho_a)-S(\rho_a||\rho_\beta(H_c)) ,
\end{equation}
where $\rho_a=\frac{1}{2}(\rho_c^{(1)}+\rho_c^{(2)})$. Under ${\cal G}_2$, we have complete control over $\frac{1}{2}(\rho^{(1)}+\rho^{(2)})$, and consequently, the relative entropy can be made null. The extractable work can thus be written as:
\begin{equation}\label{work-g}
W(\rho_i,{\cal G}_2)=\,2\,k_B T \,(\ln 2)\,\left[1-S(\frac{\rho_i^{(1)}+\rho_i^{(2)}}{2})\right],
\end{equation}
which determines an uncontrollable entropy:
\begin{equation}\label{us-g}
S_u(\rho_i,{\cal G}_2)=\,2\,S(\frac{\rho_i^{(1)}+\rho_i^{(2)}}{2})-S(\rho_i).
\end{equation}
Following the previous derivation, it is straightforward to extend these results to the case of $N$ qubits under a control set of local independent  ${\cal L}_N$ or common   ${\cal G}_N$ control sets. We write the results directly:
\begin{equation}\label{work-elen}
W(\rho_i,{\cal L}_N)=k_B T \,(\ln 2)\,\left[N-\sum_{k=1}^N S(\rho_i^{(k)})\right]
\end{equation}
\begin{equation}\label{us-elen}
S_u(\rho_i,{\cal L}_N)=-S(\rho_i)+\sum_{k=1}^N S(\rho_i^{(k)})
\end{equation}
\begin{equation}\label{work-gn}
W(\rho_i,{\cal G}_N)=\,N\,k_B T \,(\ln 2)\,\left[1-S\left(\frac{1}{N}\sum_{k=1}^N{\rho_i^{(k)}}\right)\right]
\end{equation}
\begin{equation}\label{us-gn}
S_u(\rho_i,{\cal G}_N)=\,N\,S\left(\frac{1}{N}\sum_{k=1}^N{\rho_i^{(k)}}\right)-S(\rho_i).
\end{equation}

\section{Uncontrollable entropies for $N$ distinguishable qubits under ${\cal F}_N$ control sets}
\label{sec:FN}

In this appendix we assume that a USITIR engine works with a system of $N$ distinguishable qubits, under the control set ${\cal F}_N$ defined in eq.\eqref{def:FN}. We deliberately dedicate an appendix to this problem to facilitate comparison with the bosonic case, treated in the next section. We work in the $N^2-$dimensional Hilbert space spanned by the tensor product of  the $\sigma_z^{(k)}\,,\,k=1,\ldots,N$, operator eigenstates. 

We intend to find the uncontrollable entropy $S_u(\rho_i,{\cal F}_N)$ as defined in section \ref{sec:qwork}.  Accordingly, we have to find the minimum of the relative entropy of 
$\rho_1$ with respect to $\rho_2$, where $\rho_1$ ranges over all possible states connected to $\rho_i$ via an unitary operator $U_1$ built from the control set ${\cal F}_N=\{\sum_k \sigma_z^{(k)}\}$, while $\rho_2$ may vary over the Gibbs state $\rho_2(h):=e^{F_N h}\,Z^{-1}(h)$, where $h$ is a real constant and $Z(h)=(e^h+e^{-h})^N$. The function to be minimized is
\begin{equation}\label{tbmz}
S(\rho_2||\rho_1)= - S(U_1\rho_i U_1^\dagger)-\Traza{U_1\rho_i U_1^\dagger\,\log_2\,\rho_2(h)}.
\end{equation}
The first term is the Von Neumann entropy of $\rho_i$ and does not change under unitary transformations. In addition, $U_1$ and the $\rho_2(h)$ Gibbs state commute, so that, considering the cyclicity of the trace operation, we conclude that the result does not depend on $U_1$.  Therefore, we are left with the maximization of the term given by 
\begin{equation}\label{def:mag}
J(h):=\Traza{\rho_i \,\log_2\,\rho_2(h)},
\end{equation}
on account of the factorizability of $\rho_2(h)$, it follows 
\begin{equation}\label{uma}
J(h)=\sum_{k=1}^N\Traza{\rho_i^{(k)} \,\log_2\,\rho_2^{(k)}(h)},
\end{equation}
where $\rho_2^{(k)}(h)=Z(h)^{1/N}\,e^{h\sigma_z}$ does not depend on $k$. Moreover,  $\rho_2^{(k)}(h)$ is diagonal in the computational basis, so that only the decohered  $\rho^{(k)}_{i,d}$ part of $\rho_i^{(k)}$ contributes, where $\left(\rho^{(k)}_{i,d}\right)_{m,n}:=\delta_{m,n}\,\left(\rho_i^{(k)}\right)_{m,n}$.  
With this considerations, $J(h)$ reads
\begin{equation}\label{maz}
J(h)= N \Traza{\rho_{i,d}  \,\log_2\,\rho_2^{(k)}(h)},
\end{equation}
where 
\begin{equation}\label{def:rhod}
\rho_{i,d}:= \frac{1}{N}\sum_{k=1}^N \rho^{(k)}_{i,d},
\end{equation}
which allows us to rewrite $J(h)$ as
\begin{equation}\label{mau}
J(h)= -N S(\rho_{i,d}||\rho_2^{(1)}(h))-S(\rho_{i,d}),
\end{equation}
so that the relative entropy, which is always non-negative,  can be made null. The right strategy is tuning the $h$ parameter to match the expected value of $F_N$ for the input and the Gibbs states.  The uncontrollable entropy then results
\begin{equation}\label{unod}
S_u(\rho_i,{\cal F}_N)=N S(\rho_{i,d})-S(\rho_i),
\end{equation}
and the maximum extractable work is
\begin{equation}\label{dosd}
W=k_B T\,(\ln 2)\,N\,(1-S(\rho_{i,d})).
\end{equation}

Some values of uncontrollable entropies and maximum extractable works for $N=2$ that are used in the paper are represented in Table \ref{tab:dos}. 
\begin{table}[t]
\begin{tabular}{|c|c|c|}
\toprule
\hline
\hline
$\rho_i$ & $S_u(\rho_i,{\cal F}_2)$ & $W(\rho_i,{\cal F}_2)$  \\ \hline
$\ket{00}\bra{00}$ & $0$ & $k_B \, T\,\ln 4$ \\ \hline
$\ket{10}\bra{10}$ & $k_B \, T\,\ln 4$ & $0$ \\ \hline
$\ket{01}\bra{01}$ & $k_B \, T\,\ln 4$ & $0$ \\ \hline
$\ket{11}\bra{11}$ & $0$ & $k_B \, T\,\ln 4$ 
\\
\hline\hline\bottomrule
\end{tabular}
\caption{Values for the uncontrollable entropy $S_u(\rho_i,{\cal F}_2)$ and maximum extractable work $W(\rho_i,{\cal F}_2)$ in a  distinguishable two-qubit system for different input states $\rho_i$.}\label{tab:dos}
\end{table}


\section{Uncontrollable entropies for $N$ bosonic qubits}
\label{sec:bosons}

In this appendix we assume that a USITIR engine works with a system of $N$ indistinguishable qubits, obeying bosonic statistics, under the control set ${\cal F}_N$ defined in eq.\eqref{def:FN}. Much of the exposition goes along the same lines as in appendix \ref{sec:FN}, but we prefer to repeat some considerations in order to improve readability. We work in the $(1+N)-$dimensional Hilbert space spanned by the basis ${\cal B}_N=\{\ket{0},\ket{1},\ldots,\ket{N}\}$ whose vectors represent states with well-defined values of the occupation numbers in the $\sigma_z$ operator eigenstates. That is, $\ket{n}$ is the state that corresponds to $n$ bosons in the $\sigma_z=1$ subspace and $N-n$ in the $\sigma_z=-1$. In the  ${\cal B}_N$ basis the matrix form  of the only operator $F_N$ in the control set reads
\begin{equation}\label{Fm}
F_N:=\left(
\begin{array}{cccc}
-N&0&\cdots&0\\
0&-(N-2)&\cdots&0\\
\vdots&\vdots&\vdots&\vdots\\
0&0&\cdots&N
\end{array}
\right).
\end{equation}
\begin{table}
\begin{tabular}{|c|c|c|}\toprule \hline\hline
$\rho_i$ & $S_u(\rho_i,{\cal F}_2)$ & $W(\rho_i,{\cal F}_2)$  \\ \hline
$\ket{0}\bra{0}$ & $0$ & $k_B \, T\,\ln 3$ \\ \hline
$\ket{1}\bra{1}$ & $k_B \, T\,\ln 3$ & $0$ \\ \hline
$\ket{2}\bra{2}$ & $0$ & $k_B \, T\,\ln 3$ \\ 
\hline\hline\bottomrule
\end{tabular}
\caption{Values for the uncontrollable entropy $S_u(\rho_i,{\cal F}_2)$ and maximum extractable work $W(\rho_i,{\cal F}_2)$ in two-qubit bosonic systems for different input states $\rho_i$ expressed in the occupation number basis.}\label{tab:bos}
\end{table}
Our first task is to find the uncontrollable entropy $S_u(\rho_i,{\cal F}_N)$ as defined in section \ref{sec:qwork}.  Accordingly, we have to find the minimum of the relative entropy of 
$\rho_1$ with respect to $\rho_2$, where $\rho_1$ ranges over all possible states connected to $\rho_i$ via an unitary operator $U_1$ built from ${\cal F}_N$ and $\rho_2$ may vary over the Gibbs state $e^{F_N h}\,Z^{-1}(h)$, where $h$ is a real constant. The function to be minimized is
\begin{equation}\label{tbm}
S(\rho_2||\rho_1)= - S(U_1\rho_i U_1^\dagger)-\Traza{U_1\rho_i U_1^\dagger\,\log_2\,\frac{e^{F_N h}}{Z(h)}}.
\end{equation}
The first term is the Von Neumann entropy of $\rho_i$ and can not be minimized. In addition, $U_1$ and the $\rho_2$ Gibbs state commute, so that, considering the cyclicity of the trace operation, we conclude that the result does not depend on $U_1$.  Therefore, we are left with the maximization of the term given by 
\begin{equation}\label{def:ma}
J(h):=\Traza{\rho_i \,\log_2\,\frac{e^{F_N h}}{Z(h)}},
\end{equation}
which can be easily evaluated in the ${\cal B}_N$ basis. If $\rho_{jk}$ are the the components of $\rho_i$ in ${\cal B}_N$, $J(h)$ reads
\begin{equation}\label{ma}
J(h)=-\log_2 Z(h)+\frac{1}{\ln 2}\sum_{j=1}^{N+1} \rho_{jj}F_{N,jj} h.
\end{equation}
Now we equate its derivative to zero in order to find the optimum $h^*$ value for $h$
\begin{equation}\label{maj}
\sum_{j=1}^{N+1} \rho_{jj}F_{N,jj} =\frac{Z'(h^*)}{Z(h^*)},
\end{equation}
which, in other words, states that the system must evolve unitarily until reaching a Hamiltonian for which the expected occupation numbers of the Gibbs and the input states are equal. 

As a function of $h^*$, the uncontrollable entropy is
\begin{equation}\label{ucb}
S_u(\rho_i,{\cal F}_N)=-J(h^*)-S(\rho_i),
\end{equation}
and the maximum extractable work is
\begin{equation}\label{mwb}
W=k_B T\,(\ln 2)\,(\log_2(N+1)+J(h^*)).
\end{equation}
Some values of uncontrollable entropies and maximum extractable works for $N=2$ that are used in the paper are represented in Table \ref{tab:bos}. 


\end{document}